\documentclass[dvips, 12pt]{article}
\usepackage{hyperref}
\newcommand{\ra}{\rightarrow}

\newcommand{\be}{\begin{equation}}
\newcommand{\ee}{\end{equation}}
\newcommand{\ba}{\begin{eqnarray}}
\newcommand{\ea}{\end{eqnarray}}
\newcommand{\bi}{\begin{itemize}}
\newcommand{\ei}{\end{itemize}}
\newcommand{\Tr}{{\rm Tr}}

\newcommand{\Z}{{\bf Z}}
\newcommand{\R}{{\bf R}}
\newcommand{\C}{{\bf C}}
\newcommand{\p}{\partial}

\newcommand{\CP}{{\bf P}}

\newcommand{\Ncal}{{\mathcal N}}

\newcommand{\Ocal}{{\mathcal O}}

\newcommand{\Dcal}{{\mathcal D}}

\newcommand{\Kahler}{K\"{a}hler }

\newcommand{\nn}{\nonumber}

\usepackage{amsfonts}
\usepackage{amsmath, amssymb}
\usepackage{graphicx}
\usepackage{psfrag}
\begin{document}
\baselineskip=15.5pt
\renewcommand{\theequation}{\arabic{section}.\arabic{equation}}
\pagestyle{plain} \setcounter{page}{1}
\bibliographystyle{utcaps}
\begin{titlepage}

\leftline{\tt hep-th/0409270}

\vskip -.8cm

\rightline{\small{\tt CALT-68-2522}}

\begin{center}

\vskip 3.7 cm

{\Large {\bf Derivation of Calabi-Yau Crystals from Chern-Simons Gauge Theory}}

\vskip 2.5cm
{\large Takuya Okuda}

\vskip 0.8cm

California Institute of Technology, Pasadena, CA 91125, USA

\smallskip
{\tt takuya@theory.caltech.edu}

\vskip 3.5cm

{\bf Abstract}

\end{center}
We derive new crystal melting models from Chern-Simons theory on
the three-sphere.
Via large $N$ duality,
these models compute amplitudes for A-model on the resolved conifold.
The crystal is bounded by two walls whose distance
corresponds to the \Kahler
modulus of the geometry.
An interesting phenomenon is found where  the \Kahler modulus is shifted by the presence of non-compact D-branes.
We also discuss the idea of using the crystal models as means of proving
more general large $N$ dualities to all order in $g_s$.
\end{titlepage}

\newpage


\section{Introduction}

Topological string theory \cite{Witten:1992fb,Bershadsky:1993cx} is currently undergoing
a drastic paradigm change.
Reshetikhin, Okounkov, and Vafa \cite{Okounkov:2003sp} realized that various amplitudes
for topological A model on $\C^3$ can be expressed in terms
of classical statistical models of a melting crystal.
Iqbal, Nekrasov, and Vafa \cite{Iqbal:2003ds} proposed to interpret the crystals in terms of quantum foam
or \Kahler gravity,
which is the target space theory of A-model closed string theory.
Mathematically speaking, this means that Gromov-Witten invariants are related
to the so-called Donaldson-Thomas invariants \cite{MNOP1,MNOP2}.

Central to the dramatic paradigm shift in topological string is
the interpretation of the Calabi-Yau crystal as describing the violent
fluctuations of topology and the geometry at microscopic scales.
This is reminiscent of geometric transition, where open string theory and closed string are related
via a local change of  topology and geometry.
Or rather, when crystal picture is combined with geometric transition,
one naturally expects that the geometric change is part of the gravitational fluctuations or quantum foam.
In this paper, we  realize this expectation and make it precise.

We propose a crystal melting model that describes the A model closed strings
on the resolved conifold $\Ocal(-1)\oplus \Ocal(-1)\ra \CP^1$.
Our model is a simple modification of the model for $\C^3$.
The \Kahler gravity interpretation leads one to view the gluing
prescription of the topological vertices as computing partition functions
of crystal models for general toric Calabi-Yau manifolds.
Although the resolved conifold was discussed in that context in
\cite{Iqbal:2003ds},
what we propose is different from the prescription described there.
Indeed it was emphasized in \cite{Saulina:2004da} that ``the global rule of melting is absent for closed strings on toric Calabi-Yau manifolds with more than one
fixed point of the toric action''.
One of the purposes of the present paper is to amend this situation.

Our model is obtained from the large $N$ dual Chern-Simons theory on $S^3$.
We show that the Chern-Simons theory can be formulated as  a simple unitary matrix model
that involves a theta function.
This representation is then used to obtain a free field formula for the Chern-Simons theory,
which is interpreted in terms of a statistical model.

It is also possible to introduce non-compact D-branes to the crystal,
enlarging the arena of study to include open strings.
It was shown in \cite{Saulina:2004da}  for $\C^3$
that this corresponds to having defects in the
crystal.
In Chern-Simons theory, the observables are the Wilson loops
that go around the circles in various knots and links in the three-manifold.
We show that the computation of a Wilson loop along an unknot
can be nicely done in the unitary matrix model.
This then translates to a natural crystal model with defects that represents
some number of non-compact D-branes intersecting the $\CP^1$ in the
resolved conifold.
The fact that these D-branes fit neatly into the crystal shows that our
model of the Calabi-Yau crystal is a natural one.

The crystal melting model is also a useful computational tool.
For one crystal model, there are two ways to represent it in terms
of free bosons and fermions.
We use this freedom to explicitly compute certain amplitudes in Chern-Simons
theory.
We find an interesting phenomenon where the \Kahler modulus of the resolved
conifold is shifted by a multiple of $g_s$ in the presence
of non-compact D-branes.
We also discuss the possible application of the crystal representation
to prove more general examples of topological string large $N$ duality
to all order in $g_s$.

The plan of the paper is as follows.
In section \ref{model} we propose a crystal melting model and demonstrate that it computes
the partition function for the resolved conifold.
In section \ref{open} we explain how the crystal picture naturally arises from the dual open string theory.
We also discuss the non-perturbative mismatch.
In section \ref{dbranes} we derive from Chern-Simons theory the crystal models
for non-compact D-branes, realizing them as defects in the crystal.
Section \ref{generallargeN} discusses the possible application of the crystal
computation as a way to prove large $N$ dualities to all order in $g_s$.

\section{Crystal melting model for the resolved conifold}\label{model}

Let us recall the crystal model for $\C^3$ \cite{Okounkov:2003sp}.
The zero-energy configuration is the positive octant $x,y,z\geq 0$
in $\R^3$ filled with atoms.
Here an atom at $(x_0,y_0,z_0)$ is a filled box
$\{(x_0+s_x,y_0+s_y,z_0+s_z)|0\leq s_x,s_y,s_z\leq 1\}$.
We consider removing atoms from the corner.
The allowed configurations are defined recursively as follows:
The configuration where the whole octant is filled is allowed.
If  an allowed configuration has an atom at $(x_0,y_0,z_0)$ such that there are no atoms in
the region $\{(x,y,z)|x<x_0,y<y_0,z<z_0\}$,
one can remove the atom at $(x_0,y_0,z_0)$ to obtain another allowed configuration.
The allowed configurations are also called 3D Young diagrams,
in analogy with the familiar counterpart in two dimensions.
The partition function is
\ba
Z=\sum_\pi q^{|\pi|},
\ea
where the summation is over 3D Young diagrams $\pi$,
and $q=e^{-g_s}$, $|\pi|$ is the number of atoms removed.
This partition function agrees with the partition function of A-model closed
strings on $\C^3$.
This fact can be proved by the use of free field techniques familiar in
string theory \cite{Okounkov:2003sp}.
Below we generalize the technique to the situations of our interest.

The model we propose for the resolved conifold is the following.
We add one more condition that further
restricts the allowed configurations:
Atoms in the region $x\geq N$ cannot be removed.
Here $N$ is related to the \Kahler modulus $t$ as $t=g_s N$.
Note that this condition introduces a ``wall'' that together
with the original three walls constitutes the toric diagram for the resolved conifold.

\begin{figure}\label{crystals}
(a) \scalebox{.45}{
\includegraphics{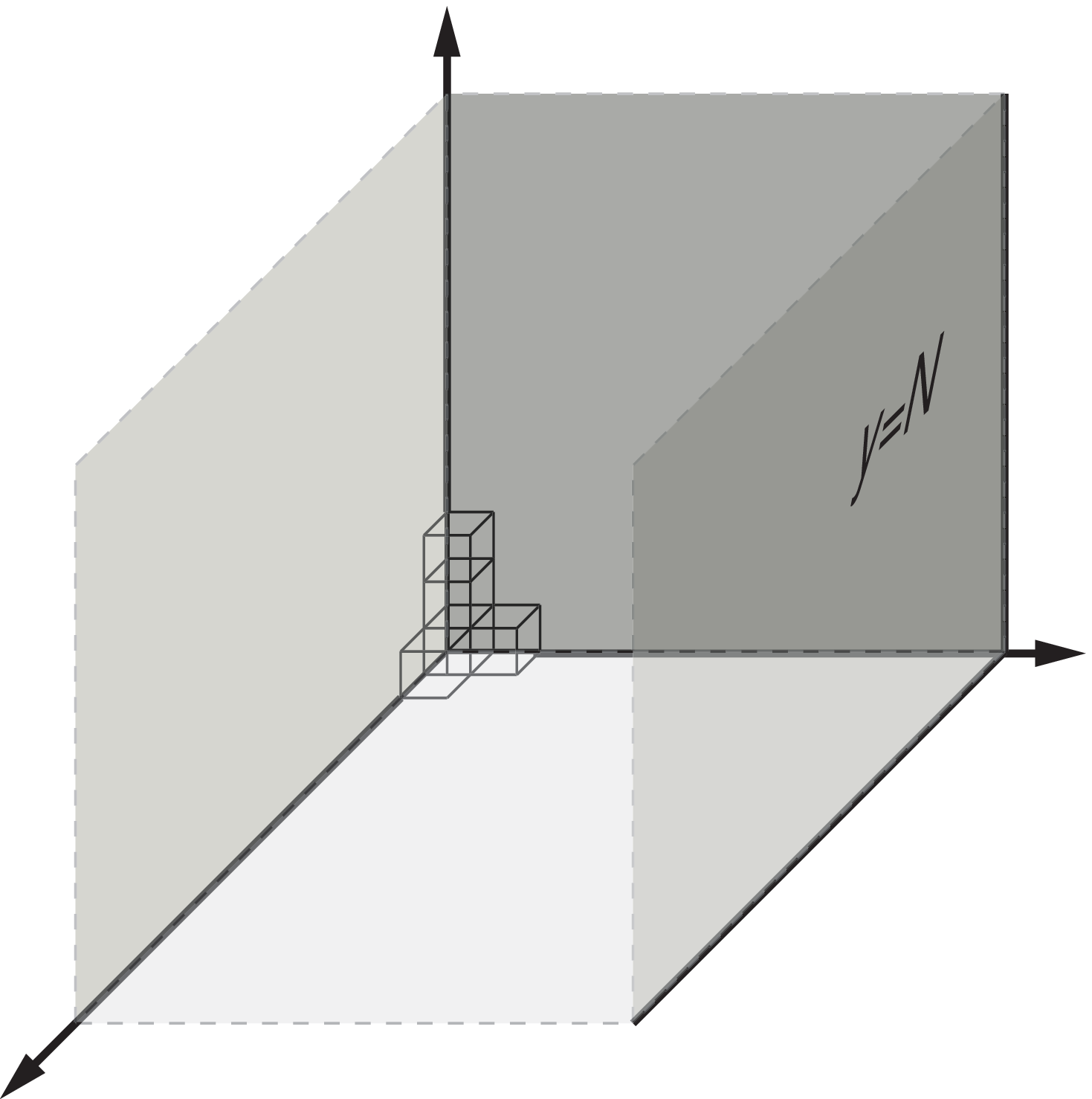}
}
(b) \scalebox{.45}{
\includegraphics{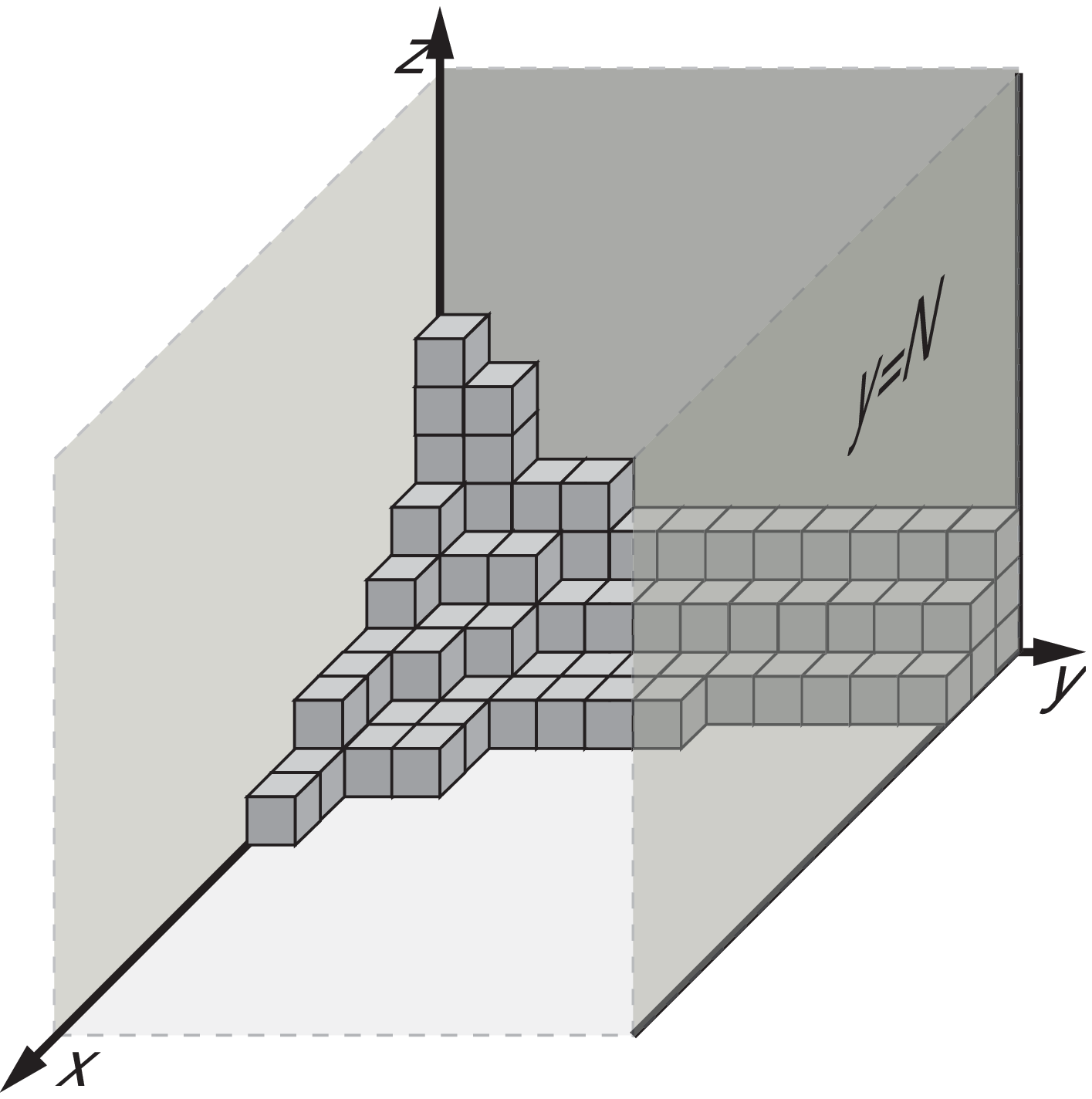}
}
\caption{(a) The crystal melting model for the resolved conifold
$\Ocal(-1)\oplus\Ocal(-1)\ra \CP^1$.
The edges, drawn as solid lines, of the positive octant bounded by the wall at $y=N$ form the toric diagram of the resolved conifold.
(b)
Many atoms have been removed from the crystal.
Atoms cannot be removed from the region beyond the wall at $y=N$.
}
\end{figure}

Now we demonstrate that this crystal melting program indeed reproduces
the partition function for the resolved conifold.
For this purpose, we express the partition function in terms of free fermions and bosons:
\ba
&&\psi(z)=\sum_{r\in \Z+1/2} \frac{\psi_r}{z^{r+1/2}},~
\bar{\psi}(z)=\sum_r \frac{\bar{\psi}_r}{z^{r+1/2}},\\
&&\{\psi_r,\bar{\psi}_s\}=\delta_{r+s,0},\\
&&
\phi(z)=x_0-i\alpha_0\log z+i\sum_{n\neq 0}\frac{\alpha_n}{nz^n},\\
&&\left[\alpha_{m},\alpha_{n}\right]=m\delta_{m+n,0}.
\ea
These are related via
\ba
&&i\p \phi(z)=:\psi\bar{\psi}(z):,~~\psi(z)=:e^{i\phi(z)}:,~~\bar{\psi}(z)=:e^{-i\phi(z)}:.
\ea
Now we define
\ba
\Gamma_\pm(z)=\exp\sum_{n>0}\frac{z^{\pm n}}{n}\alpha_{\pm n}.
\ea
It is well known that neutral (zero momentum in the bosonic language) fermionic Fock states
are labelled by (2D) Young diagrams $\mu$ which we denote as
$\mu=(\mu_1\geq \mu_2...\geq \mu_d>0)$.
More explicitly, such Fock states are given by
\ba
|\mu\rangle&=&\prod_{i=1}^\infty \psi_{i-\mu_i-1/2}|0\rangle\rangle\nn\\
&=&
\prod_{i=1}^d \bar{\psi}_{-a_i}\psi_{-b_i}|0\rangle,\label{statemu}
\ea
where $|0\rangle\rangle$ is the state that is annihilated by all $\bar{\psi}_r, r\in \Z+1/2$, and we have defined
\ba
a_i=\mu_i-i+1/2,~~b_i=\mu^t_i-i+1/2.
\ea
$\mu^t$ is the transposed Young diagram.
The Virasoro zero mode $L_0$ counts the number $|\mu|$ of boxes in the Young diagram $\mu$:
\ba
L_0|\mu\rangle=|\mu||\mu\rangle.
\ea

\begin{figure}[tbp]
(a)
\scalebox{.45}{\includegraphics{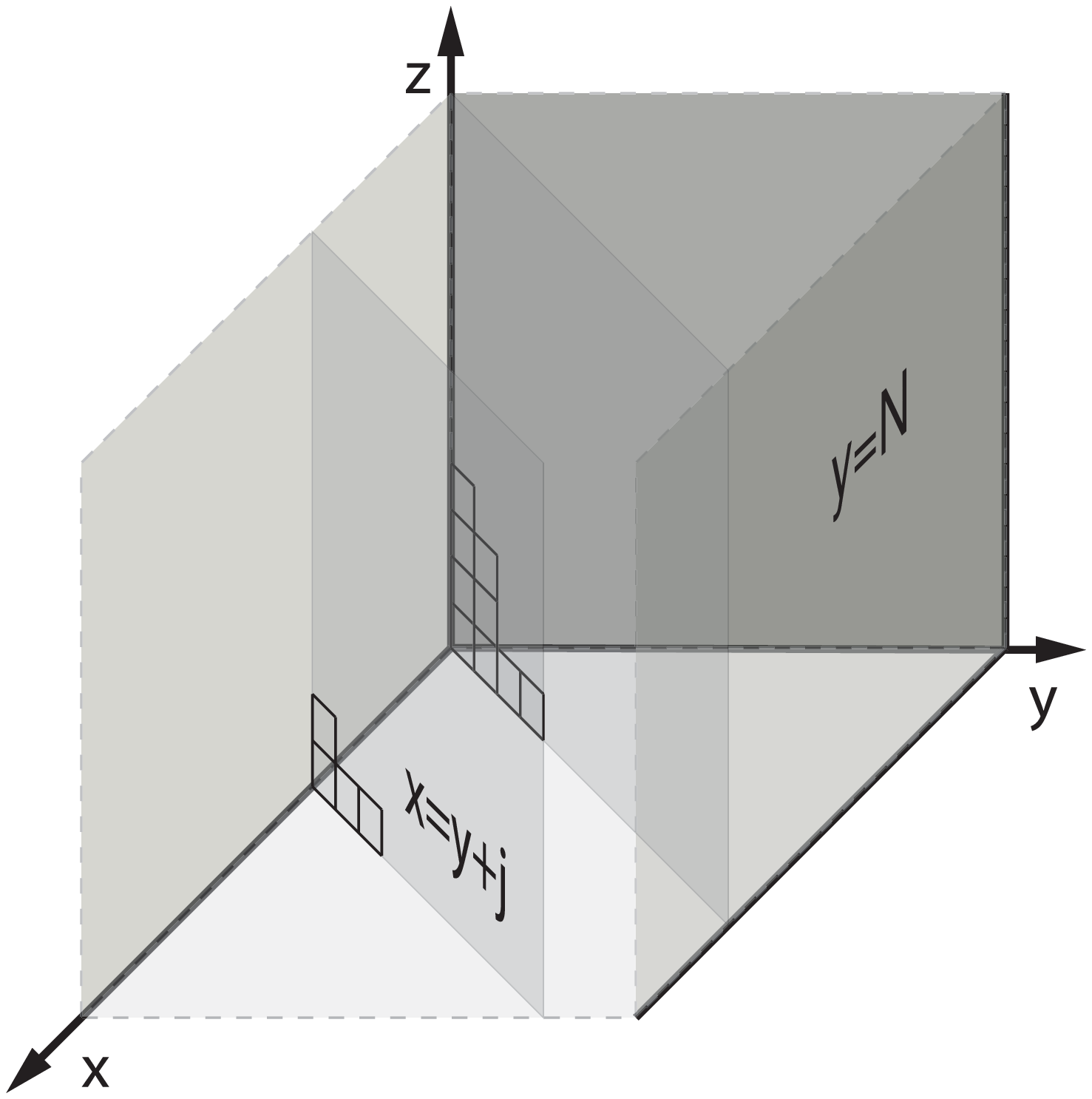}}
(b)
\scalebox{.45}{\includegraphics{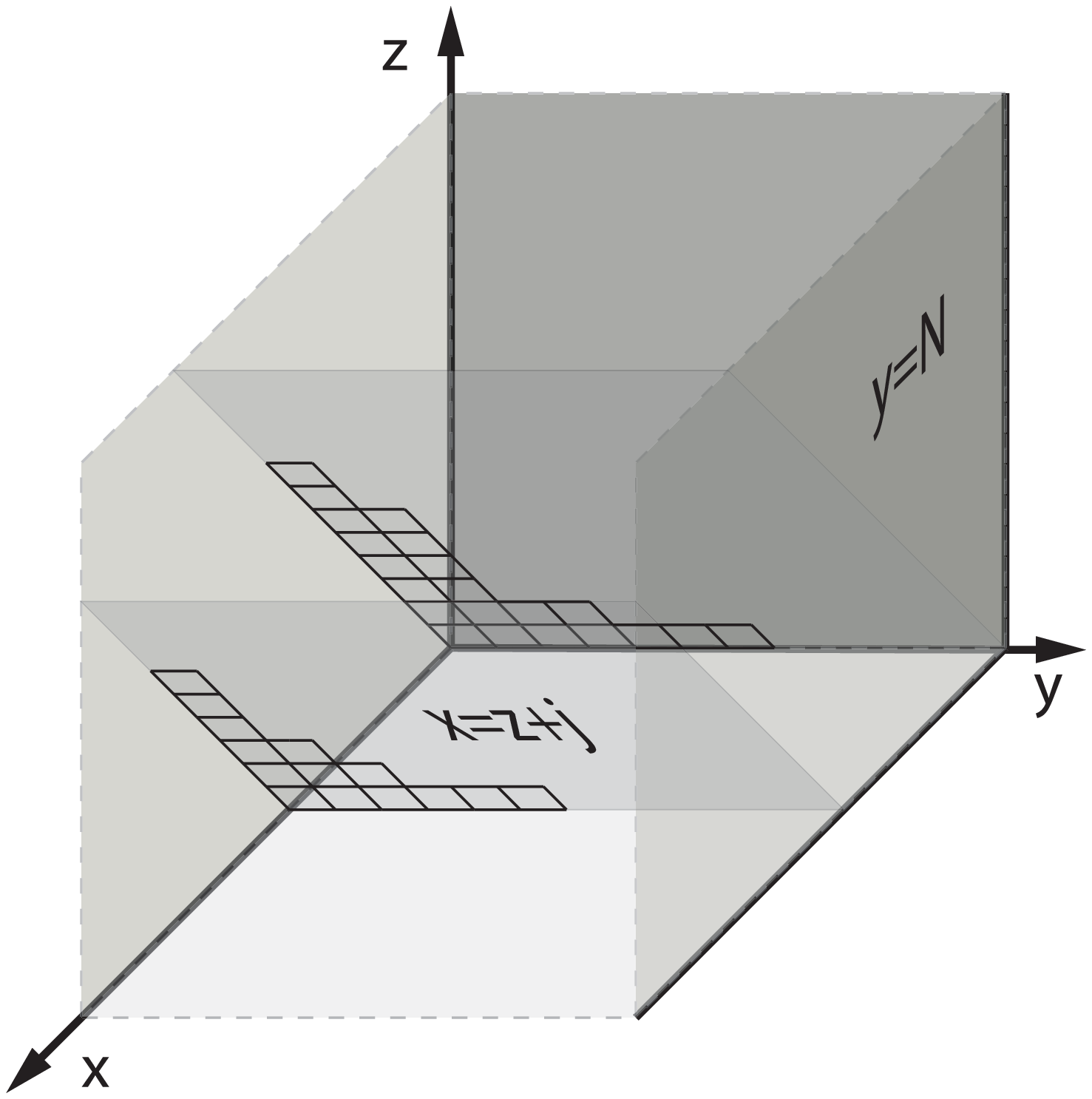}}
\caption{
(a)Closed string slicing:
This slicing by planes $x=y+j$ allows one to compute the closed string amplitude as in eq. (\ref{closedstringslicing}).
(b)Open string slicing: Another slicing of the crystal by planes $x=z+j$ corresponds to
the free field representation eq. (\ref{openstringslicing}) obtained from Chern-Simons theory.}
\label{crystalslicing}
\end{figure}

Two Young diagrams $\lambda$ and $\mu$ are said to interlace (and we write $\lambda \succ \mu$)
if they satisfy
\ba
\lambda_1\geq \mu_1\geq \lambda_2\geq\mu_2\geq...
\ea
In other words, $\lambda$ and $\mu$ interlace if and only if $\lambda$ contains $\mu$ and
$\mu$ contains $\lambda$ with the first row removed.
The interlacing condition is equivalent to the local condition for
two Young diagrams one finds by slicing the allowed configuration
of the crystal by the planes $x=y+j$ and $x=y+j+1$ \cite{Okounkov:2003sp}.
The operators $\Gamma_{\pm}(z)$ are useful because of the properties
\ba
\Gamma_+(1)|\lambda\rangle=\sum_{\lambda\succ\mu}|\mu\rangle,\nn\\
\Gamma_-(1)|\lambda\rangle=\sum_{\mu\succ\lambda}|\mu\rangle.
\ea
The partition function for the crystal can be written as
\ba
&&Z_{\rm crystal}(q,t=g_s N)=
\langle 0|\left(\prod_{n=1}^\infty
q^{L_0}\Gamma_+(1)\right)
q^{L_0}
\left(
\prod_{m=1}^N
\Gamma_-(1)q^{L_0}\right)|0\rangle
\nn\\
&=&
\langle 0|\prod_{n=1}^\infty\Gamma_+(q^{n-1/2})\prod_{m=1}^N\Gamma_-(q^{-(m-1/2)})|0\rangle.
\ea
This can be understood as slicing the crystal by planes $x=y+j, j\in \Z$.
Note that we have a finite product of vertex operators acting on $|0\rangle$.
This restricts a 3D Young diagram to have a trivial 2D Young diagram
on the slice $x=y-N$.
The interlacing conditions then imply that
the 2D Young diagrams must have at most one row
on the slice $x=y-N+1$,
two rows on $x=y-N+2$, etc.
Thus the free field correlator represents
a crystal model bounded by a wall at $y=N$.
See figure \ref{crystalslicing}(a).

Now we can explicitly compute the partition function.
\ba
Z_{\rm crystal}(q,t=g_s N)
&=&\langle 0|e^{-\sum_{n>0}\frac{\alpha_n}{n[n]}}e^{-\sum_{n>0}\frac{1-q^{Nn}}{n[n]}\alpha_{-n}}
|0\rangle\nn\\
&=&e^{\sum_{n>0}\frac{1-q^{Nn}}{n[n]^2}}\nn\\
&=&M(q)e^{-\sum_{n>0}\frac{e^{-nt}}{n[n]^2}}.\label{closedstringslicing}
\ea
Taking $N\ra \infty$ pushes the wall at $y=N$ to infinity,
and the partition function reduces to the result for $\C^3$.

This crystal model and the resulting amplitude are different from those
discussed in \cite{Iqbal:2003ds}.
While our crystal has a fixed finite size in the $y$ direction,
in \cite{Iqbal:2003ds} the distance between the two crystal corners
are not fixed because two finite size 3D partitions are connected
through a region of length $t=g_s N$.
Consequently, in stead of a single power of $M(q)$ in our model,
the model \cite{Iqbal:2003ds} gives the square of $M(q)$.
More generally, a closed string partition function contains $M(q)^{\chi(X)/2}$,
where $\chi(X)$ is the Euler characteristic of the target space $X$.
If the target space $X$ is non-compact, the definition of the Euler characteristic is  ambiguous.
In the context of large $N$ duality, it is known \cite{Gopakumar:1998ki}
that one should assign the value 2 to the Euler characteristic of the resolved conifold
as we just did.
This is natural in the sense that the target space admits one \Kahler deformation
but no complex structure deformation, and the general formula for a (compact) Calabi-Yau manifold
is $\chi(X)=2\left[(\# {\rm K\ddot{a}hler~ deformations})-(\# {\rm Complex~ structure ~ deformations})\right]$.

\section{Large $N$ dual open string theory }\label{open}
In this section, we study the crystal melting problem from the point of view
of large $N$ duality.

The crystal model in the previous section can also be expressed
as
\ba
&&Z_{\rm crystal}(q,t=g_s N)\nn\\
&=&\langle 0|\prod_{n=1}^\infty\Gamma_+(q^{n-1/2}){\bf 1}_{d^t\leq N}\prod_{m=1}^\infty\Gamma_-(q^{-(m-1/2)})|0\rangle \label{openstringslicing}
\ea
Here ${\bf 1}_{d^t\leq N}$ is the operator that projects onto the
subspace spanned by $|\mu\rangle$ such that the Young diagram $\mu$
has at most $N$ columns.
This free field expression corresponds to slicing the crystal by planes
$z-x=j, j\in \Z$.
See figure \ref{crystalslicing}(b).

We  call this the ``open string slicing'' because, as we will see below, this representation of the crystal naturally arises
from Chern-Simons theory.

\subsection{Unitary matrix model for Chern-Simons theory}\label{unitarymatrixmodel}
The large $N$ duality of Gopakumar and Vafa relates $U(N)$ Chern-Simons theory on $S^3$
to topological closed string on the resolved conifold $\Ocal(-1)\oplus \Ocal(-1)\ra \CP^1$.
The dictionary is that the \Kahler modulus $t$ of the
closed string theory geometry is identified with the 't hooft parameter
$g_s N$.
Certain amplitudes on the resolved conifold,
including the closed string amplitudes,
can be computed within the framework of Chern-Simons theory.
We  now develop a unitary matrix model formulation of Chern-Simons theory,
which will be used to derive the crystal model for the resolved conifold later.

The partition function of the $U(N)$ Chern-Simons theory on $S^3$
is given by \cite{Witten:1988hf}
\ba
Z_{\rm CS}(N,k, U(N))&=&\frac{1}{(k+N)^{N/2}}\prod_{\alpha>0}2\sin\frac{\pi \alpha\cdot \rho}{k+N}.
\ea
Here $2\pi i/(k+N)=g_s$ is the string coupling constant, and the product is
over the positive roots of $SU(N)\subset U(N)$ which are given by $\alpha_{ij}=e_i-e_j\in
\C^N$($\simeq$ Cartan subalgebra) for $i<j$.
$\rho=(1/2)\sum_{\alpha>0}\alpha=\sum_{i=1}^N(\frac{N+1}{2}-i)e_i$
is the Weyl vector.
Now note the following  formula by Weyl
for the denominator of the Lie algebra characters
\ba
\prod_{\alpha>0}2\sinh(\alpha\cdot u)=\sum_{w\in W}\epsilon(w)e^{w(\rho)\cdot u},
\label{Weylden}
\ea
where $W$ is the Weyl group isomorphic to the permutation group $S_N$  and $u$ is an arbitrary element of the Cartan subalgebra of $U(N)$.
We use this identity to rewrite $Z_{\rm CS}(N,k, U(N))=:Z_{\rm CS}(g_s, N)$ as
\ba
Z_{\rm CS}(g_s, N)&=&\frac{e^{-\frac{N(N-1)\pi i}{4}}}{(k+N)^{N/2}}\sum_{w\in W} \epsilon(w)
e^{g_s w(\rho)\cdot\rho}\nn\\
&=&
\left(\frac{g_s}{2\pi}\right)^{N/2}e^{-\frac{\pi i}{4}N^2}q^{-\frac{N(N-1)}{12}}
\tilde{Z}_{\rm CS}(g_s,N)
\ea
Here we have factored out the non-trivial part of the partition function:
\ba
\tilde{Z}_{\rm CS}(g_s,N)=
\sum_{w\in W} \epsilon(w)
q^{\frac{1}{2} (w(\rho)-\rho)^2}.
\ea
$q$ is again $e^{-g_s}$.
In what follows, we ``analytically continue'' in $g_s$ and regard $g_s$ as a complex parameter
with a positive real part.
We can introduce another sum over the Weyl group and an integral
over the maximal torus as follows:\footnote{
Here we use the identity
$q^{m^2/2}=\int_0^{2\pi}\frac{d\theta}{2\pi}\vartheta_{00}(e^{i\theta};q)e^{im\theta}$.
If we instead use \\
$q^{m^2/2}=\int_{-\infty}^\infty \frac{du}{\sqrt{2\pi g_s}}e^{-\frac{u^2}{2g_s}}e^{mu}$,
we get the matrix model with
a non-compact integration region
introduced in  \cite{Marino:2002fk, Aganagic:2002wv}.
The matrix model there can  related be transformed to our unitary matrix model
via $u=i(\theta+2\pi n)$, performing the sum over $n\in \Z$ and a modular transformation.
The author thanks Hirosi Ooguri for pointing this out.
}
\ba
&&
\tilde{Z}_{\rm CS}(g_s,N)\nn\\
&=&\frac{1}{|W|}\sum_{w,w'\in W} \epsilon(w)
\epsilon(w')
q^{\frac{1}{2} (w(\rho)-w'(\rho))^2}\nn\\
&=&\frac{1}{|W|}
\int
\prod_{i=1}^N \left(\frac{d\theta_i}{2\pi}\vartheta_{00}(e^{i\theta_i};q)\right)
\sum_{w,w'\in W}
 \epsilon(w)
\epsilon(w')
e^{i (w(\rho)-w'(\rho))\cdot \theta}.
\ea
Here
\ba
\vartheta_{00}(e^{i\theta};q):=\sum_{m\in\Z}q^{\frac{m^2}{2}}e^{im\theta}
\ea
is one of Jacobi's theta functions.

By making use of the Weyl denominator formula eq. (\ref{Weylden}) again,
we get
\ba
\tilde{Z}_{\rm CS}&=&\frac{1}{|W|}
\int
\left(\prod_{i=1}^N \frac{d\theta_i}{2\pi}\vartheta_{00}(e^{i\theta_i})\right)
\left(\prod_{\alpha>0}2\sin\frac{\alpha\cdot\theta}{2}\right)^2.
\ea
The second factor now represents the Haar measure for $U(N)$ pushed down
to the maximal torus.
The partition
function can be written in a very simple form
\ba
\tilde{Z}_{\rm CS}=\int_{U(N)} dU\det\vartheta_{00}(U;q),
\ea
where the measure is normalized so that the volume of $U(N)$ is unity.
This expression holds for any gauge group of the Chern-Simons theory on $S^3$
when the corresponding Haar measure is used.

\subsection{Crystal from Chern-Simons theory}
Now we use the product formula for the theta function
\ba
\vartheta_{00}(e^{i\theta};q)&=&
\prod_{j=1}^\infty(1-q^j)(1+e^{i\theta}q^{j-1/2})(1+e^{-i\theta}q^{j-1/2})
\nn\\
&=&
\left(\prod_{j=1}^\infty(1-q^j)\right) \exp\left[
\sum_{n>0}(-1)^n\frac{e^{in\theta}+e^{-in\theta}}{n[n]}
\right]
\ea
to write
\ba
\tilde{Z}_{\rm CS}&=&\left(\prod_{j=1}^\infty(1-q^j)\right)^N
\int dU\exp\left[
\sum_{n>0}(-1)^n\frac{\Tr U^n+\Tr U^{-n}}{n[n]}
\right]\nn\\
&=&\left(\prod_{j=1}^\infty(1-q^j)\right)^N
\int dU\exp\left[
\sum_{n>0}\frac{\Tr U^n+\Tr U^{-n}}{n[n]}
\right]
\ea

To obtain the free-field expression for the
partition function,
we introduce the coherent states:
\ba
|U\rangle:=\exp\left[\sum_{n>0}\frac{1}{n}\Tr ~ U^n \alpha_{-n}\right]|0\rangle.
\ea
These states satisfy
\ba
\alpha_n |U\rangle=\Tr~ U^n|U\rangle,\\
\int dU |U\rangle\langle U|={\bf 1}_{d\leq N},
\ea
where ${\bf 1}_{d\leq N}$ is the projection to the subspace
spanned by $|\mu\rangle$ such that the number of rows in $\mu$ is less than
or equal to $N$.
This formalism was extensively used in the context of 2D Yang-Mills theory
which has recently been attracting some attention \cite{Vafa:2004qa,deHaro:2004id,deHaro:2004uz}. See \cite{Cordes:1994fc}
and the references therein.

By making use of $|U\rangle$, we can write
\ba
\tilde{Z}_{\rm CS}&=&\left(\prod_{j=1}^\infty(1-q^j)\right)^N\langle 0|
e^{\sum_{n>0}\frac{\alpha_n}{n[n]}}\int dU|U\rangle\langle U|
e^{\sum_{n>0}\frac{\alpha_{-n}}{n[n]}}|0\rangle\nn\\
&=&\left(\prod_{j=1}^\infty(1-q^j)\right)^N\langle 0|
e^{\sum_{n>0}\frac{\alpha_n}{n[n]}}
{\bf 1}_{d\leq N}e^{\sum_{n>0}\frac{\alpha_{-n}}{n[n]}}|0\rangle
\ea
Since $\alpha_n\ra -\alpha_n$ is equivalent to $R\ra R^t$ (c.f. eq. (\ref{statemu})),
we finally obtain
\ba
\tilde{Z}_{\rm CS}&=&\left(\prod_{j=1}^\infty(1-q^j)\right)^N\langle 0|
e^{-\sum_{n>0}\frac{\alpha_n}{n[n]}}
{\bf 1}_{d^t\leq N}e^{-\sum_{n>0}\frac{\alpha_{-n}}{n[n]}}|0\rangle\nn\\
&=&
\xi(q)^{-N}
Z_{\rm crystal}(q;t=g_s N)
.\label{openfree}
\ea
We have demonstrated that the open string slicing eq. (\ref{openstringslicing})
naturally arises from Chern-Simons theory.
Note that there is a mismatch by the factor $\xi(q)^{-N}$
between $\tilde{Z}_{\rm CS}$ and $Z_{\rm crystal}$,
where $\xi(q)=1/\prod_{j=1}^\infty(1-q^j)$ is the
``renormalization factor'' which was found in \cite{Saulina:2004da}
to be associated with a non-compact D-brane.
As in \cite{Saulina:2004da}, we use the modular property
of $\eta(q)=q^{1/24}\xi(q)^{-1}$, namely
$\eta(q)=\sqrt{2\pi/g_s}\eta(\tilde{q}), \tilde{q}=e^{-4\pi^2/g_s}$,
to argue that it does not
contribute to the perturbative amplitudes at genus no less than $2$
when comparing the open
and closed string sides.
For genus amplitudes, the mismatch is absorbed into the
usual ambiguities.

\section{Adding D-branes}\label{dbranes}

We can add non-compact D-branes to the system.
In the language of Chern-Simons gauge theory, this corresponds to
placing Wilson lines going through circles of links.
In the case of an unknot, we will be able to see the connection to the
description in \cite{Saulina:2004da}.

On the open string side, we consider placing a stack of $M$ non-compact D-branes in $T^\ast S^3$
intersecting the $S^3$ along an unknot $S^1$ \cite{Ooguri:1999bv}.
Since the new D-branes are non-compact, we treat them as non-dynamical,
acting as a source to the gauge fields on $S^3$ via an interaction.
This interaction is obtained by integrating out the  degrees freedom coming
from the open strings stretching between the compact D-branes wrapping the $S^3$ and
the non-compact D-branes.
Let $U\in U(N)$ and $V\in U(M)$ be the holonomies along the unknot for the gauge fields on the compact
and the non-compact D-branes, respectively.
Then the interaction can be represented as
\ba
\int \Dcal A e^{-S_{CS}[A]+\sum_{n=1}^\infty\frac{1}{n}\Tr U^n \Tr V^n}
=Z_{CS}(S^3)\langle e^{\sum_{n=1}^\infty\frac{1}{n}\Tr U^n \Tr V^n}\rangle.
\ea
The expectation value can be expanded, with the help of Frobenius' formula, as
\ba
\langle e^{\sum_{n=1}^\infty\frac{1}{n}\Tr U^n \Tr V^n}\rangle&=&
\sum_\mu \langle \Tr_\mu U\rangle \Tr_\mu V. \label{afterFrobenius}
\ea
Here $\Tr_\mu$ denotes the trace in the representation of $U(N)$ or $U(M)$ specified by the Young diagram $\mu$.

It is natural to expect that $\langle \Tr_\mu U\rangle$ in eq. (\ref{afterFrobenius}) is computed by the unitary matrix model in subsection \ref{unitarymatrixmodel}
by inserting $\Tr_\mu U$.
We now show that this is indeed correct, however
with the subtlety that the Wilson line and hence the non-compact D-branes have non-canonical framing.

The object we would like to compute is
\ba
\int dU \det \vartheta_{00}(U;q)\Tr_\mu U.
\ea
Going back to the eigenvalue integral, this is
\ba
\int\prod_{i=1}^N \frac{d\theta_i}{2\pi}\vartheta_{00}(e^{i\theta_i})
\det\left[(e^{i\theta_j})^{N-i}\right]
\det\left[(e^{-i\theta_j})^{N-i}\right]
\frac{\det\left[(e^{i\theta_j})^{\mu_i+N-i}\right]}{\det\left[(e^{i\theta_j})^{N-i}\right]},
\ea
where we have used the Jacobi-Trudy formula
$\Tr_\mu{\rm diag}(x_1,...,x_N)\equiv s_\mu(x_1,...,x_N)=\det x_j^{\mu_i+N-i}/\det x_j^{N-i}$
for the Schur polynomial\footnote{A good reference
on symmetric functions and the group theory relevant to us
is \cite{FultonHarris}.}.
After cancelling factors between the numerator and the denominator,
and performing the integrals the matrix integral reduces to
\ba
&&\frac{1}{N!}\sum_{\sigma,\sigma'\in S_N} {\rm sgn}\sigma ~{\rm sgn}\sigma'
\prod_{j=1}^N q^{\frac{1}{2}\left(\mu_{\sigma(j)}-\sigma(j)+\sigma'(j)\right)^2}\nn\\
&=&\sum_{\sigma\in S_N} {\rm sgn}\sigma
\prod_{j=1}^N q^{\frac{1}{2}\left(\mu_j-j+\sigma(j)\right)^2}\nn\\
&=&
\det\left[
q^{\frac{1}{2}(\mu_i-i+j)^2}\right].
\ea
Up to $\mu$-independent factors, this equals
\ba
q^{\frac{1}{2}\sum_{i=1}^N \mu_i(\mu_i-2i+N+1)}
\det\left[q^{(j-\frac{N+1}{2})(\mu_i-i+N)}\right].
\ea
The power of $q$ can be written as $q^{(\kappa_\mu+N|\mu|)/2}$, where
$\kappa_\mu=2\sum_{(i,j)\in \mu}(i-j)=\sum_i \mu_i(\mu_i-2i+1)$.
This is the factor one obtains when the framing of the Wilson loop is
shifted by one unit \cite{Witten:1988hf}.
The determinant is of the form that appears in the numerator of the
Jacobi-Trudy formula.
Hence we have shown that
\ba
&&\frac{\int dU \det \vartheta_{00}(U;q) \Tr_\mu U}{\int dU \det \vartheta_{00}(U;q)}\nn\\
&=&q^{(\kappa_\mu+N|\mu|)/2}
\Tr_\mu{\rm diag}(q^{-\frac{N-1}{2}},q^{-\frac{N-3}{2}},...,q^{\frac{N-1}{2}}).
\ea
Relative to the result for the canonically framed unknot \cite{Ooguri:1999bv},
we see that the matrix model computes amplitudes in the framing shifted by one unit.

\begin{figure}[tbp]
\centering
\scalebox{.45}{
\includegraphics{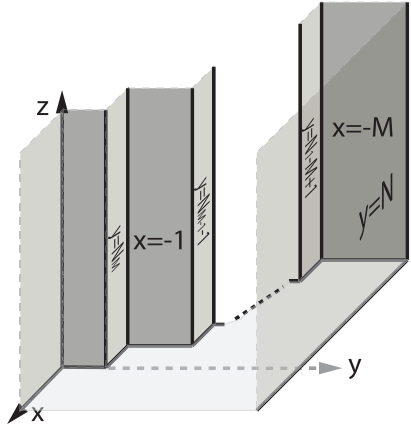}
}
\caption{The initial configuration of the crystal with defects representing multiple
non-compact D-branes intersecting $\CP^1$ in the resolved conifold.
The defects introduce faces at $y=\mu^t_1=N_1-M+1,\mu^t_2=N_2-M+2,...,\mu_{M-1}^t=N_{M-1}-1,\mu_M^t=N_M$.
}
\label{manydbranes}
\end{figure}

This vacuum expectation value of the Wilson loop can be
represented as a crystal melting model as follows.

\ba
&&\int dU \det\vartheta_{00}(U;q)\Tr_\mu U\nn\\
&=&
\xi(q)^{-N}\langle 0|e^{\sum_{n>0}\frac{\alpha_n}{n[n]}}
\int dU \Tr_\mu U|U\rangle\langle U|
e^{\sum_{n>0}\frac{\alpha_{-n}}{n[n]}}|0\rangle\nn\\
&=&
\xi(q)^{-N}\langle 0|e^{\sum_{n>0}\frac{\alpha_n}{n[n]}}
\int dU \sum_{\vec{k}}\frac{1}{z_{\vec{k}}}\chi_\mu(C(\vec{k}))\prod_{j=1}^\infty \alpha_j^{k_j}|U\rangle\langle U|
e^{\sum_{n>0}\frac{\alpha_{-n}}{n[n]}}|0\rangle,
\ea
where $\chi_\mu$ is the $S_{|\mu|}$ character of the representation specified by
$\mu$, $\vec{k}=(k_1,k_2,...)$ is an infinite vector with non-negative integer
components, and $C(\vec{k})$ is the conjugacy class of $S_{|\mu|}$ specified by
$\vec{k}$.
Now the powers of $\alpha_j$ can be moved to the left to act on $\langle 0|$.
This yields
\ba
&&\int dU \det\vartheta_{00}(U;q)\Tr_\mu U\nn\\
&=&
\xi(q)^{-N}\langle \mu|e^{\sum_{n>0}\frac{\alpha_n}{n[n]}}
\int dU |U\rangle\langle U|
e^{\sum_{n>0}\frac{\alpha_{-n}}{n[n]}}|0\rangle\nn\\
&=&
\xi(q)^{-N}\langle \mu|e^{\sum_{n>0}\frac{\alpha_n}{n[n]}}
{\bf 1}_{d\leq N}
e^{\sum_{n>0}\frac{\alpha_{-n}}{n[n]}}|0\rangle\nn\\
&=&
\xi(q)^{-N}\langle \mu^t|e^{-\sum_{n>0}\frac{\alpha_n}{n[n]}}
{\bf 1}_{d^t\leq N}
e^{-\sum_{n>0}\frac{\alpha_{-n}}{n[n]}}|0\rangle\nn\\
&=&
\xi(q)^{-N}
\langle \mu^t|\prod_{n=1}^\infty \Gamma_+(q^{n-1/2})
{\bf 1}_{d^t\leq N}
\prod_{m=1}^\infty
\Gamma_-(q^{-(m-1/2)})|0\rangle\nn\\
&=&\xi(q)^{-N} q^{\sum_{i=1}^\infty (i-1/2)\mu^t_i} Z_{\rm crystal}^{\rm D-branes},
\ea
where we have defined
\ba
Z_{\rm crystal}^{\rm D-branes}:= q^{-\sum_{i=1}^\infty (i-1)\mu^t_i}
\langle \mu^t|\prod_{n=1}^\infty \Gamma_+(q^{n})
{\bf 1}_{d^t\leq N}
\prod_{m=1}^\infty\Gamma_-(q^{-(m-1)})|0\rangle
\ea
This free field correlator together with the power of $q$ represents, in the open string slicing, the partition function of the
crystal melting model whose initial configuration is shown in figure \ref{manydbranes}.
The power of $q$ ensures that the initial configuration has zero energy.

It is possible to express the multi-D-brane crystal in the closed string slicing,
which is a slight generalization of the free field representation in \cite{Saulina:2004da}.
\ba
&&Z_{\rm crystal}^{\rm D-branes}\nn\\
&=&\langle 0|\prod_{n=1}^\infty \Gamma_+(q^{n-1/2})\prod_{m=1}^{N_M}\Gamma_-(q^{-(m-1/2)})\Gamma_+(q^{-(N_M+1/2)})\nn\\
&&~\times
\prod_{m=N_M+2}^{N_{M-1}}\Gamma_-(q^{-(m-1/2)})\Gamma_+(q^{-(N_{M-1}+1/2)})
\prod_{m=N_{M-1}+2}^{N_{M-2}}\Gamma_-(q^{-(m-1/2)})\Gamma_+(q^{-(N_{M-2}+1/2)})...\nn\\
&&~\times
\prod_{m=N_2+2}^{N_1}\Gamma_-(q^{-(m-1/2)})\Gamma_+(q^{-(N_1+1/2)})
\prod_{m=N_1+2}^{N+M}\Gamma_-(q^{-(m-1/2)})|0\rangle.
\ea
Here $\mu^t_1=N_1-M+1,\mu^t_2=N_2-M+2,...,\mu_{M-1}^t=N_{M-1}-1,\mu_M^t=N_M$.
In the closed string slicing, it is possible to explicitly evaluate the correlator to write it as a product.
This also provides us with an interpretation of $N_i$ as positions of D-branes
and exhibits an interesting shift in the \Kahler modulus:

\ba
&&Z_{\rm crystal}^{\rm D-branes}\nn\\
&=&
\left(\prod_{n=1}^\infty
\prod_{
  1\leq m\leq N+M,
  m\neq N_j+1
}
\frac{1}{1-q^{n+m-1}}\right)
\prod_{i=1}^M \prod_{
N_i+2\leq m\leq N+M,
 m\neq N_j+1}
\frac{1}{1-q^{m-N_i-1}}\nn\\
&=&
\xi(q)^M \left[\prod_{1\leq i<j\leq M}(1-e^{a_j-a_i})\right]
\nn\\
&&~~\times
M(q)
\exp\left(-\sum_{n=1}^\infty\frac{e^{-n\tilde{t}}}{n[n]^2}\right)
\prod_{i=1}^M \exp\left(
\sum_{n=1}^\infty \frac{e^{-na_i}+e^{-n(\tilde{t}-a_i)}}{n[n]}
\right).\label{multidbraneproductform}
\ea
Here we have defined $a_i:=g_s(N_i+1/2), ~i=1,...,M$ and $\tilde{t}:=g_s(N+M)=t+g_s M$.
Again, $\xi(q)$ can be essentially ignored
in the perturbative computation due to the modular property of $\eta(q)=q^{1/24}\xi(q)^{-1}$.
The second factor $\prod_{i<j}(1-e^{a_j-a_i})$ is also present in the multi-brane case of
\cite{Saulina:2004da}, and written in this way is independent of $g_s$.
This is the amplitude for $M$ non-compact D-branes in the resolved conifold, which can
be defined as the \Kahler quotient
\ba
\{(X_I)\in \C^4:|X_1|^2+|X_2|^2-|X_3|^2-|X_4|^2={\rm Re}~\tilde{t}\}/U(1)
\ea
with $U(1)$ action by charges $(1,1,-1,-1)$.
The geometry of the D-branes is \cite{Aganagic:2001nx}
\ba
|X_1|^2-{\rm Re}(a_i)=|X_2|^2-{\rm Re}~(\tilde{t}-a_i)=|X_3|^2=|X_4|^2,~~
\sum_I \arg X_I=0.
\ea
One thing that is interesting in our computation is that the \Kahler parameter is shifted
from $t=g_s N$ to $\tilde{t}=t+g_s M$.
It has been known (see, for example, \cite{Aganagic:2003db}) that the presence of D-branes can shift the effective size of the geometry
by the string coupling times the number of D-branes.
Here we have found another such phenomenon.
The genus zero part of eq. (\ref{multidbraneproductform})
in the case of a single D-brane agrees with the results in \cite{Aganagic:2000gs}.

The fact that that non-compact D-branes can be nicely incorporated to
the crystal confirms that our crystal model of the resolved conifold
is a natural one.

\section{More general large $N$ dualities, instanton counting,
and geometric engineering}\label{generallargeN}
So far we have been discussing the Calabi-Yau crystal in the context of Gopakumar-Vafa duality
($T^\ast S^3\Leftrightarrow \Ocal(-1)\oplus\Ocal(-1)\ra \CP^1)$,
the simplest example of large $N$ duality in topological string theory.
There is a family of generalizations of the large $N$ duality which is worth considering
in relation to Calabi-Yau crystal.
The example of Gopakumar and Vafa
is simple enough to prove the duality (at least at the level
of free energies and some open string amplitudes) by direct calculations.
However, our derivation of the resolved conifold crystal from Chern-Simons theory
can be viewed as a complicated way of proving the duality.
In this section we discuss the possible application of the ideas in the present paper
to prove more general large $N$ dualities.

Aganagic, Klemm, Marino, and Vafa made a conjecture in \cite{Aganagic:2002wv} that that the duality of Gopakumar and Vafa
still holds after taking a $\Z_n$ orbifold on the both sides of duality.
On the closed string side, this produces $A$-type topological closed string theory
living on the particular fibration of the $A_{n-1}$ ALE space over $\CP^1$.
The geometry has $n$ \Kahler moduli, the sizes of the base $\CP^1$ and $n$ additional $\CP^1$ that blow up
the $A_{n-1}$ singularity.
On the open string side, we again get Chern-Simons theory, this time living on the lens space $L(n,1)\simeq S^3/\Z_n$.
Also after taking the orbifold, the relevant open string theory is a sector of Chern-Simons theory
which contains one classical solution.
A classical solution can be specified by a holonomy $\exp[2\pi i/N{\rm diag}(\stackrel{N_1}{\overbrace{1,...,1}},\stackrel{N_2}{\overbrace{2,...,2}}
,...,
\stackrel{N_n}{\overbrace{n,...,n}})]$ along the generator of the homotopy group.
The \Kahler parameters are then to be identified with linear combinations of the t' Hooft parameters $g_s N_i, i=1,...,n$.

The $n=2$ duality was tested via perturbative computations by the authors
who proposed the duality \cite{Aganagic:2002wv}.
For general $n$ and a related duality, checks have been done by showing that the matrix models describing the sector of Chern-Simons theory
leads to the spectral curves which are the non-trivial parts of the Calabi-Yau manifolds mirror to the A-model closed string geometries
\cite{Halmagyi:2003ze,Halmagyi:2003mm,Yasnov:2004yd}.
The worldsheet derivation of the Gopakumar-Vafa duality \cite{Ooguri:2002gx}
has also been generalized for these large $N$ dualities \cite{Okuda:2004rg}.

There are $n+1$ choices ($m=0,1,...,n+1$ in the notation of \cite{Iqbal:2003zz}) one can make when one fibers the $A_{n-1}$ ALE space over $\CP^1$.
The closed string geometry that is dual to the $S^3/\Z_n$ Chern-Simons theory
is precisely the fibration $m=0$ \cite{Halmagyi:2003mm} that was shown to correspond to Nekrasov's
instanton counting \cite{Nekrasov:2002qd,Nekrasov:2003rj} for the 5D $SU(n)$ gauge theory with vanishing Chern-Simons term
\cite{Iqbal:2003ix,Iqbal:2003ds}.
(For the correspondence with non-zero Chern-Simons term, see \cite{Tachikawa:2004ur}.)
Nekrasov's correspondence between topological closed strings and 5D gauge theory
has been discussed in \cite{Eguchi:2003sj,Zhou0311,Eguchi:2003it} by making use of the topological vertex
\cite{Aganagic:2003db}.
As discussed in the introduction, the computation via the topological vertex
is closely related to the Calabi-Yau crystal.
In particular, the computation takes the form of an expansion in $q=e^{-g_s}$.

As we saw in the previous section,
the Chern-Simons theory also naturally leads to an expansion in $q$.
Hence, it is plausible that one will be able to prove the generalized large
$N$ dualities to all order in $g_s$ by proving that the partition functions are the same
on the both sides as functions of $q$ \cite{Okuda}.

Let us also make an observation on the appearance of the unitary matrix model.
In \cite{Dijkgraaf:2002vw}, the question of finding matrix models
that compute the Seiberg-Witten solutions of $\Ncal=2$ gauge theories
was addressed.
The matrix models in \cite{Aganagic:2002wv} can be regarded
as computing amplitudes in the 5D gauge theories with the same number
of supercharges.
By taking a double scaling limit,
which is the familiar field theory limit of geometric engineering
\cite{Katz:1996fh},
one can compute amplitudes for 4D $\Ncal=2$ gauge theories from these
matrix models.
By using the technique in this paper, it is possible to
rewrite the matrix models in \cite{Aganagic:2002wv} as unitary matrix
models.
These are similar to, and can be regarded as generalizations of,
the unitary matrix model (Gross-Witten one plaquette model \cite{Gross:1980he}) that was considered in \cite{Dijkgraaf:2002vw} for
the $SU(2)$ gauge theory.

\section*{Acknowledgments}
I would like to thank Jaume Gomis, Hikaru Kawai, Takeshi Morita,
and Sanefumi Moriyama for useful discussions. I am also grateful
to Hirosi Ooguri for giving helpful comments and reading the
manuscript. This research is supported in part by DOE grant
DE-FG03-92-ER40701.

\bibliography{bib5}

\end{document}